\documentclass{article}
\usepackage{multicol}
\usepackage[english]{babel}
\usepackage{blindtext}
\usepackage[utf8]{inputenc}
\usepackage{amssymb}
\usepackage{graphicx}
\usepackage{dblfloatfix} 
\graphicspath{ {images/} }
\usepackage{amsmath}
\usepackage{mathtools}
\usepackage[a4paper,left=2.1cm ,right=2.1cm , top=2.0cm , bottom=2.0cm ]{geometry} 
\usepackage{url}
\usepackage[export]{adjustbox}
\usepackage{mathtools}

\DeclarePairedDelimiter\floor{\lfloor}{\rfloor}
\usepackage{caption}
\usepackage{subcaption}
\usepackage{color}
\usepackage{array}
\usepackage{calc}
\usepackage{mwe}
\usepackage{fancyhdr}
\newcommand{\beqn}{\begin{eqnarray}}
\newcommand{\eeqn}{\end{eqnarray}}

\begin{document}

\begin{center}  
\textbf{Sudden appearance of a thick Dirichlet wall in a cavity}\\
\end{center}

\begin{center}  
\textbf{Saad TAIL}\\
\end{center}
\begin{center}  
\textbf{\texttt{\emph{Faculty of sciences, Aix-Marseille Université, 13007 Marseille, France}}}\\
\end{center}

\begin{center}
\noindent\rule{15cm}{0.4pt} 
\end{center}
\begin{abstract}
 \hspace{5mm}We investigate the vacuum properties of a massless scalar field theory in constrained spatial geometry, namely,  instantaneous appearance of a thick Dirichlet boundary  inside a one dimensional (1D) Dirichlet cavity and divides it into two parts. Our work presents the calculations for the energy density and particle number density created. The expression of the energy density is found to be dependent on the size nature of the appearing wall, more precisely, it depends on whether the length of the wall is rational or irrational.
\end{abstract}
\begin{center}
\noindent\rule{15cm}{0.4pt} 
\end{center}


\begin{multicols}{2}
  
 \section{Introduction}
    \hspace{5mm}The idea of the emptiness of vacuum has been shown to be untrue by the prediction of what is known as the quantum fluctuations, meaning, the fluctuations of the quantum fields are always existing in space.  One of the consequences of these fluctuations is an effect called the Casimir effect (named after Hendrik Casimir who, in 1948, predicted their existence \cite{H.Cazimir}).  The Casimir effect implies that the presence of physical (material) objects affects  virtual  (vacuum)  fluctuations  nearby.   The  total  energy  of  vacuum  fluctuations  depends  on  shapes, orientations, and distances between physical bodies so that the bodies experience mutual forces related to the minimization of the vacuum energy. These fluctuations manifest themselves in the form of virtual particles which spontaneously appear and disappear in short periods of time allowed by the Heisenberg uncertainty principle.\\
    \hspace{5mm}We know that in the dynamical Casimir effect (DCE) see \cite{DCE}, in which one (or two) of the walls of a cavity undergoes an acceleration, particles are created, see \cite{Moore70, Fulling} for details, and this is due to the excitation of the quantum fields by the dynamics (motion) of the wall. In addition to DCE, a sudden change in the geometry of a cavity also excites the field vacuum confined inside this cavity and the phenomenon of particle creation is retrieved.\\ 
    \hspace{5mm}In this paper, we consider a test scalar field in a one-dimensional (1D) Dirichlet cavity. We assume that a two-sided thick Dirichlet wall can appear instantaneously at the center of the cavity, which mimics the sudden change in the spacetime topology.\\
    \hspace{5mm}This study is important for instance in studying the quantum fields reaction to the case of a topology change in spacetime due to singularities and wormholes creation. The formation of a wormhole can happen in a disconnection of space (as seen in Fig. \ref{fig:split}).
    Those phenomena engender significant excitation of the vacuum fields in spactime giving rise to particle creation. Particle creation in some situations analogous to spacetime splitting (Fig. \ref{fig:split}) has been investigated in the literature, particularly by Anderson and DeWitt \cite{DeWitt86}.\\
    The paper consists of two section 2 and 3. Section 2 is divided into two subsections in which we discuss two issues of the same nature, namely, the length-type of the thick Dirichlet wall. In section 2.1, we look at the irrational quantity length of the wall while the rational quantity is the subject of section 2.2. Section 3 is devoted to a conclusion and a small discussion. Finally, Three long and detailed calculations of some equations in the paper are given in the appendices A and B.\\
    
    \includegraphics[width=8cm]{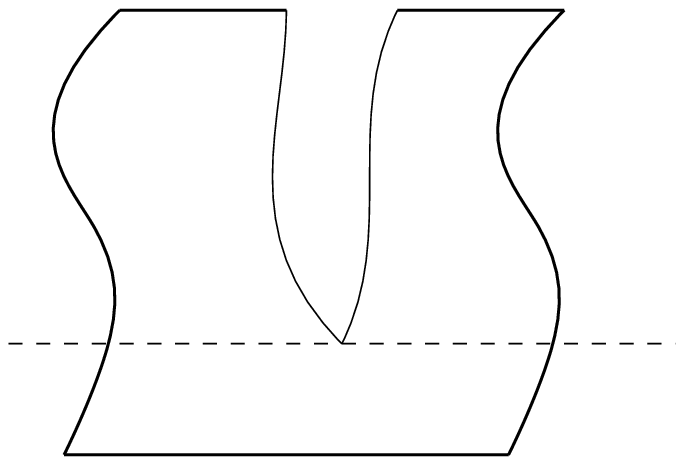}
    Figure 1: a schematic drawing of spacetime splitting.
    \label{fig:split}

 \section{Sudden appearance of thick Dirichlet wall}
    \hspace{5mm}This section generalizes the discussion of the sudden appearance of thin Dirichlet wall effect in \cite{Artcl-2016} to a thick wall. Fig. \ref{fig:thkwall} shows the setup of the problem. The discussion and calculation here will be similar to the ones used in \cite{Artcl-2016}. In our configuration we consider a (1D) Dirichlet cavity of size $L$, for which a massless scalar field, call it $\phi$, vanishes at the boundaries of the cavity i.e. $\phi(t,-L/2)=\phi(t,L/2)=0$. At $t=0$, a thick Dirichlet wall of size $\epsilon$, appears instantaneously in the middle of the cavity (see Fig. \ref{fig:thkwall}). Now, what we need to do is to express the field before and after the appearance of that two-sided boundary in the cavity. For this purpose, we solve the Klein-Gordon (KG) equation
    \begin{equation}\label{eq:1}
        (\partial_{t}^2 - \partial_{z}^2)\phi  = 0 
    \end{equation}
    with $-\infty<t<\infty$ and $-L/2<z<L/2$.\\
    For $t<0$, we have the following set of solutions

    \includegraphics[width=8cm]{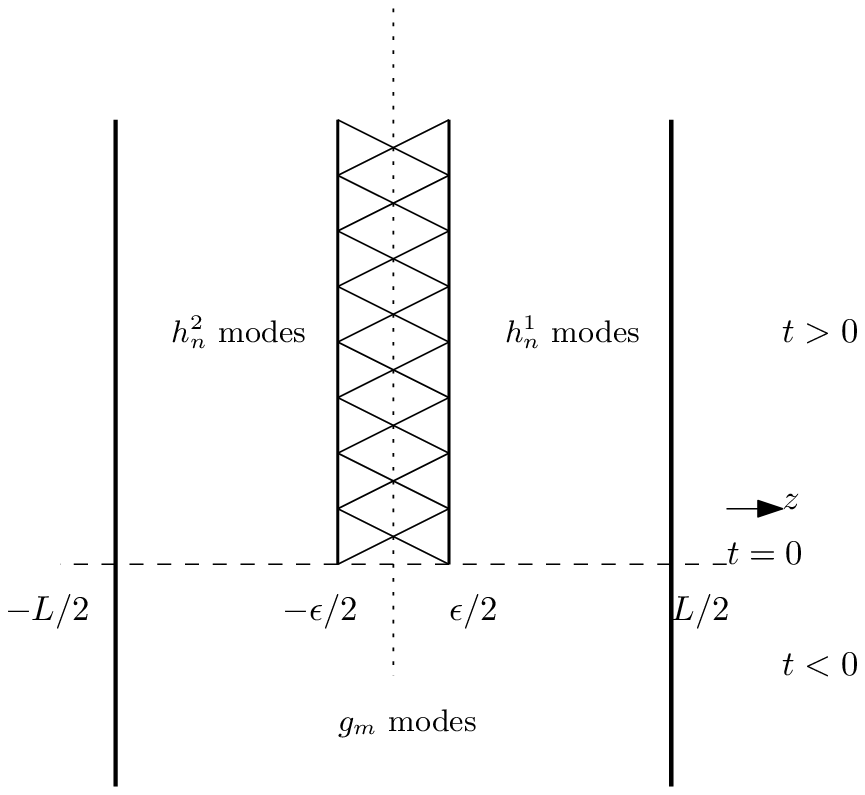}
    \label{fig:thkwall}
    Figure 2: Instantaneous appearing of thick wall in a cavity (the dashed area stands for no field inside)\\

    \begin{equation} \label{eq:64}
    g_m = \frac{1}{\sqrt{m\pi}}e^{-i q_m t} 
    \begin{cases}
      & \cos(q_m z) \text{\hspace{0.5cm};   m odd} \\
      & \sin(q_m z) \text{\hspace{0.5cm};   m even} \\
    \end{cases} 
    \end{equation}
    with $ q_m=\frac{m\pi}{L}$. The $g_{m}$ functions also satisfy the orthogonality conditions defined by
    \begin{equation}\label{ortho-condi}
    \langle g_{m}, g_{m'} \rangle=-\langle g_{m}^{*}, g_{m'}^{*} \rangle =\delta_{mm'} \textbf{\hspace{5mm}}\langle g_{m}, g_{m'}^{*} \rangle=0
    \end{equation}
    The product in (\ref{ortho-condi}) is called the KG product defined as $ \langle f, g \rangle := i\int_0^{L} ( f^{*}\partial_{t}g-\partial_{t}f^*g) dz$ in an interval $[0,L]$.
    The field then is put in its quantized form as
    \begin{equation}\label{eq:one}
    \phi=\sum_{m}^{\infty}(b_m g_m + b_{m}^{\dagger}g_{m}^*)
    \end{equation}    
    where $b_{m}^{\dagger}$ and $b_m$ are the creation and annihilation operators, respectively.\\
    
    For $t>0$, (meaning after the appearance of the wall) the solution of (\ref{eq:1}) gives the following set of solutions

    \begin{equation} \label{eq:88}
    h_{n}^{1} =
    \begin{cases}
      & 0  \text{\hspace{4cm}; $-\frac{L}{2}<z<\frac{-\epsilon}{2}$}\\
      & \pm \frac{1}{\sqrt{n\pi}}e^{-i p_n t} \sin(p_n (z-\epsilon/2))\text{\hspace{0.1cm}; $\frac{\epsilon}{2}<z<\frac{L}{2}$} \\
    \end{cases} 
    \end{equation}
    and
    \begin{equation} \label{eq:89}
    h_{n}^{2} =
    \begin{cases}
      & \mp \frac{1}{\sqrt{n\pi}}e^{-i p_n t} \sin(p_n (z+\epsilon/2))\text{\hspace{0cm}; $-\frac{L}{2}<z<\frac{-\epsilon}{2}$} \\
      & 0  \text{\hspace{4.2cm}; $\frac{\epsilon}{2}<z<\frac{L}{2}$}\\      
    \end{cases} 
    \end{equation}
    where $p_n=\frac{2n\pi}{L-\epsilon}$ and the index "$l$" stands for the two regions of the cavity after the appearance of the boundary in the middle (Fig. \ref{fig:thkwall}). The $h_{n}^{l}$ functions also satisfy the orthogonality conditions as
    \begin{equation}
    \langle h_{n}^{l}, h_{n'}^{l'} \rangle=-\langle h_{n}^{l*}, h_{n'}^{l'*} \rangle =\delta_{nn'}\delta_{ll'}, \textbf{\hspace{1.5mm}}\langle h_{n}^{l}, h_{n'}^{l'*} \rangle=0
    \end{equation}
    Again, in this case, the field is put in its quantized form as \begin{equation}\label{eq:two}
        \phi=\sum_{l=1}^{2}\sum_{n=1}^{\infty}(a_{n}^l h_{n}^{l} + a_{n}^{l\dagger}h_{n}^{l*})
    \end{equation}

    with $a_{n}^{l\dagger}$ and $a_{n}$ being the creation and annihilation operators. By Bogoliubov transformation, we can write the following

    \begin{equation}\label{eq:68} 
    a_{n}^l=\sum_{m=1}^{\infty}(\alpha_{mn}^l b_{m} + \beta_{mn}^{l*}b_{m}^{\dagger})
    \end{equation}

    where $\alpha_{mn}$ and $\beta_{mn}$ are the Bogoliubov coefficients determined by $\langle h_{n}^l,g_m \rangle$ and $-\langle h_{n}^{l*},g_m \rangle$, respectively. The creation and annihilation operators satisfy the commutation relations, namely

    \begin{equation}\label{eq:69}  
    [b_m,b_{m'}^{\dagger}]=\delta_{mm'} \hspace{2cm} [b_m,b_{m'}]=0
    \end{equation}
    \begin{equation}\label{eq:70}  
    [a_{n}^l,a_{n'}^{l'\dagger}]=\delta_{ll'}\delta_{nn'} \hspace{1.8cm} [a_{n}^{l},a_{n'}^{l'}]=0
    \end{equation}

    Replacing (\ref{eq:68}) in (\ref{eq:70}) and using  (\ref{eq:69}), we get that $\alpha_{mn}$ and $\beta_{mn}$ verify

    \begin{equation}\label{appxA1}
    \sum_{m=1}^{\infty}(\alpha_{mn}^l\alpha_{mn'}^{l'*}-\beta_{mn}^{l*}\beta_{mn}^{l'})=\delta_{ll'}\delta_{nn}
    \end{equation}

    \begin{equation}\label{appxA2}
    \sum_{m=1}^{\infty}(\alpha_{mn}^l\beta_{mn'}^{l'*}-\beta_{mn}^{l*}\alpha_{mn'}^{l'})=0
    \end{equation}
    When we compute the Bogoliubov coefficients using our expressions of $h_{n}^{l}$ and $g_{m}$, we find the following 

    \begin{equation}\label{eq:90}
    \alpha_{mn}^{l} =
    \begin{cases}
      & \frac{1}{\pi \sqrt{mn}}\frac{2n}{2n-m(1-\epsilon/L)}\cos(\frac{m\pi}{2}\frac{\epsilon}{L})\text{ \hspace{0.1cm}; $m$ odd($\neq$)} \\
      & (-1)^l\sqrt{\frac{n}{m}}\sin((m-2n)\frac{\pi}{2})\text{ \hspace{0.6cm}; m odd(=)} \\
      & \frac{1}{\pi \sqrt{mn}}\frac{2n}{2n-m(1-\epsilon/L)}\sin(\frac{m\pi}{2}\frac{\epsilon}{L})\text{ \hspace{0.1cm}; m even($\neq$)} \\
      & (-1)^{l-1}\sqrt{\frac{n}{m}}\cos((m-2n)\frac{\pi}{2})\text{ \hspace{0.1cm}; m even(=)} \\
    \end{cases} 
    \end{equation}

    \begin{equation}\label{eq:91}
    \beta_{mn}^{l} =
    \begin{cases}
      & \frac{1}{\pi\sqrt{mn}}\frac{2n}{2n+m(1-\epsilon/L)}\cos(\frac{m\pi}{2}\frac{\epsilon}{L}) \text{ \hspace{0.1cm}; m odd($\neq$)} \\
      & 0 \text{ \hspace{4.2cm}; m odd(=)} \\      
      & \frac{1}{\pi\sqrt{mn}}\frac{2n}{2n+m(1-\epsilon/L)}\sin(\frac{m\pi}{2}\frac{\epsilon}{L}) \text{ \hspace{0.1cm}; m even($\neq$)} \\ 
      & 0 \text{ \hspace{4.2cm}; m even(=)} \\   
    \end{cases} 
    \end{equation}
    where, in order to avoid the overwriting in the expressions, I referred to the condition of "$m$ odd and $\frac{\epsilon}{L}\neq 1 - 2 \frac{n}{m}$" as "odd($\neq$)" and for "$m$ odd and $\frac{\epsilon}{L}= 1 - 2 \frac{n}{m}$" as "odd(=)". The same thing applied for "$m$ even". In the last case, since $\epsilon/L$ is non-negative, we must have $ m\geq2n$ indeed. The details on how these conditions arise are found in appendix \ref{AppendixA}\\

    Notice that for $\epsilon=0$, the above expressions reduce to those derived in \cite{Artcl-2016}.\\

    The quantity of interest is $\langle 0_g|T_{tt}|0_g\rangle$ for $t<0$ and for $t>0$. The subscript ”$g$” in $|0_g\rangle$ stands for the field vacuum state when we have only the $g_m$ modes

    for $t <0$, one would expect to get the well know result: $-\frac{\pi}{24L^2}$, and of course it is the case as it is shown here
    
    \begin{equation}\label{eqQq5133}
\begin{split}
    & \langle 0_g| T_{tt} |0_g\rangle \\
    & = \langle 0_g| \frac{1}{2}[(\partial_t \phi)^2 + (\partial_z \phi)^2] |0_g\rangle \\
    & = \frac{1}{2}\sum_{m=1}^{\infty} (|\partial_t g_m|^2 +|\partial_z g_m|^2)\\
    & = \frac{1}{2}\sum_{\stackrel{{}_{\mathrm{odd}}}{m = 1}}^{\infty} (|\partial_t g_m|^2 +|\partial_z g_m|^2) + \frac{1}{2}\sum_{\stackrel{{}_{\mathrm{even}}}{m = 2}}^{\infty} (|\partial_t g_m|^2 +|\partial_z g_m|^2)\\
    & = \frac{1}{2}\sum_{\stackrel{{}_{\mathrm{odd}}}{m = 1}}^{\infty}\frac{q_{m}^2 }{m\pi} + \frac{1}{2}\sum_{\stackrel{{}_{\mathrm{even}}}{m = 2}}^{\infty}\frac{q_{m}^2}{m\pi}\\
    & = \frac{1}{2}\sum_{m=1}^{\infty}\frac{m \pi}{L^2} \xrightarrow{\text{regularized}}-\frac{\pi}{24L^2}
\end{split}
\end{equation}

For $t>0$, the computation is more complicated and in order to be able to compute this quantity in a simplified manner, we introduce new coordinates $x_\pm$ defined as

\begin{equation}\label{eq:76}
    t= \frac{1}{2}(x_{+} - x_{-})
\end{equation}
\begin{equation}\label{eq:77}
    z= \frac{1}{2}(x_{+} + x_{-})
\end{equation}

Now, $T_{tt}$ could be written as 
\begin{equation}\label{eqT_tt}
    T_{tt}=T_{++}+T_{--}
\end{equation}
\hspace{5mm}and
\begin{equation}\label{appA}
    T_{\pm \pm}= (\partial_\pm \phi)^2
\end{equation}
    where the subscripts "$+$" and "$-$" mean "$x_+$" and "$x_-$", respectively. Replacing equation (\ref{eq:two}) into (\ref{appA}) and then computing the vacuum expectation value gives the equation (\ref{eq:92}) below

\end{multicols}

\beqn
\langle 0_g| T_{\pm\pm} |0_g\rangle & = & \sum_{n=1}^{\infty}\sum_{n'=1}^{\infty}\sum_{l=1}^{2} 
\bigg( \sum_{\stackrel{{}_{\mathrm{odd}(\neq)}}{m = 1}}^{\infty} \bigl[(\alpha_{mn}^{l}\beta_{mn'}^{l}+\beta_{mn}^{l}\alpha_{mn'}^{l}) \Re(\partial_{\pm}h_{n}^{l}\partial_{\pm}h_{n'}^{l}) \nonumber \\ 
& & +(\alpha_{mn}^{l}\alpha_{mn'}^{l}+\beta_{mn}^{l}\beta_{mn'}^{l})\Re(\partial_{\pm}h_{n}^{l}\partial_{\pm}h_{n'}^{l*}) \bigr] + \sum_{\stackrel{{}_{\mathrm{odd}(=)}}{m=3\hspace{1mm}(  m\geq2n)}}^{\infty}[(\alpha_{mn}^{l}\alpha_{mn'}^{l})\Re(\partial_{\pm}h_{n}^{l}\partial_{\pm}h_{n'}^{l*})]\bigg) \nonumber \\
& & + \sum_{n=1}^{\infty}\sum_{n'=1}^{\infty}\sum_{l=1}^{2} 
\bigg(\sum_{\stackrel{{}_{\mathrm{even}(\neq)}}{m = 2}}^{\infty} 
 [(\alpha_{mn}^{l}\beta_{mn'}^{l}+\beta_{mn}^{l}\alpha_{mn'}^{l})\Re(\partial_{\pm}h_{n}^{l}\partial_{\pm}h_{n'}^{l}) \nonumber \\ 
& & +(\alpha_{mn}^{l}\alpha_{mn'}^{l}+\beta_{mn}^{l}\beta_{mn'}^{l})\Re(\partial_{\pm}h_{n}^{l}\partial_{\pm}h_{n'}^{l*})] + \sum_{\stackrel{{}_{\mathrm{even}(=)}}{m=2\hspace{1mm}(  m\geq2n)}}^{\infty}[(\alpha_{mn}^{l}\alpha_{mn'}^{l})\Re(\partial_{\pm}h_{n}^{l}\partial_{\pm}h_{n'}^{l*})]\bigg)\,.
\label{eq:92}
\eeqn\\

\begin{multicols}{2}
\hspace{5mm}Again, it is worth noticing that for $\epsilon=0$, the above expression coincides with the its peer in \cite{Artcl-2016}, namely, equation (2.23). As it is seen from the expression of Bogoliubov coefficients (equations (\ref{eq:90}) and (\ref{eq:91})), we compute the energy density for different value-type of the ratio $\epsilon/L$, namely irrational and rational.
\subsection{Case 1: $\epsilon/L$ irrational}
We assume that $\epsilon/L$ is irrational, this makes the  "\emph{not equal} ($\neq$)" condition always satisfied, and the terms of the "\emph{equal} ($=$)" are, therefore, omitted. Hence, by considering this and replacing in (\ref{eq:92}) by the explicit expressions, we get

\end{multicols}
  
 \begin{equation}\label{eq:93}
\begin{split}
\langle 0_g| T_{\pm\pm} |0_g\rangle =
& \frac{4}{\pi (L-\epsilon)^2}\sum_{\stackrel{{}_{\mathrm{odd}}}{m = 1}}^{\infty}\bigg(  \cos^2(\frac{m\pi}{2}\frac{\epsilon}{L})\frac{1}{m}\bigg[w^2\sum_{n=1}^{\infty}\cos(\frac{2n\pi}{L-\epsilon}(x_{\pm}-\frac{\epsilon}{2}))  +m^2\sum_{n=1}^{\infty}\frac{\cos(\frac{2n\pi}{L-\epsilon}(x_{\pm}-\frac{\epsilon}{2}))}{n^2 -(m/w)^2} \bigg]^2\frac{1}{w^4}\\
& + \cos^2(\frac{m\pi}{2}\frac{\epsilon}{L})m \bigg[ \sum_{n=1}^{\infty}\frac{n\sin(\frac{2n\pi}{L-\epsilon}(x_{\pm}-\frac{\epsilon}{2}))}{n^2 -(m/w)^2} \bigg]^2 \frac{1}{w^2} \bigg)+\\
& \frac{4}{\pi (L-\epsilon)^2}\sum_{\stackrel{{}_{\mathrm{even}}}{m = 2}}^{\infty}\bigg(  \sin^2(\frac{m\pi}{2}\frac{\epsilon}{L})\frac{1}{m}\bigg[w^2\sum_{n=1}^{\infty}\cos(\frac{2n\pi}{L-\epsilon}(x_{\pm}-\frac{\epsilon}{2})) 
 +m^2\sum_{n=1}^{\infty}\frac{\cos(\frac{2n\pi}{L-\epsilon}(x_{\pm}-\frac{\epsilon}{2}))}{n^2 -(m/w)^2} \bigg]^2\frac{1}{w^4}\\
& + \sin^2(\frac{m\pi}{2}\frac{\epsilon}{L})m \bigg[ \sum_{n=1}^{\infty}\frac{n\sin(\frac{2n\pi}{L-\epsilon}(x_{\pm}-\frac{\epsilon}{2}))}{n^2 -(m/w)^2} \bigg]^2 \frac{1}{w^2} \bigg)
\end{split}
\end{equation} 
  
\begin{multicols}{2} 
  where $w=\frac{2}{1-\frac{\epsilon}{L}}$. Notice that this expression is periodic with respect to $v:=x_{\pm}-\frac{\epsilon}{2}$with a period of $L-\epsilon$ and invariant under translation $v\rightarrow -v$, hence, we can compute the quantity in the interval $0\leq v < L-\epsilon $. We can represent a schematic of our system by the diagram in Fig. 3. \ref{fig:fff} The sums in the first and the third lines of equation (\ref{eq:93}) are computed by making use of the following formulae (see \cite{Artcl-2019,japan})
 
\begin{equation}\label{eq:94}
\sum_{k=1}^{\infty}\cos(\frac{2k\pi}{a}y)= -\frac{1}{2}+\frac{a}{2}\sum_{l=-\infty}^{\infty}\delta(y-la)
\end{equation}
valid for all $y$. And
\begin{equation}\label{eq:95}
\sum_{k=1}^{\infty}\frac{\cos(ky)}{k^2-a^2}=-\frac{\pi}{2a}\cos[a(\pi-y)]\csc(a\pi)+\frac{1}{2a^2}
\end{equation}
valid for $0\leq y \leq 2\pi$. The sums inside the  square brackets in the first and the third line become
\end{multicols}
  
    \begin{center}
     \includegraphics[width=13cm]{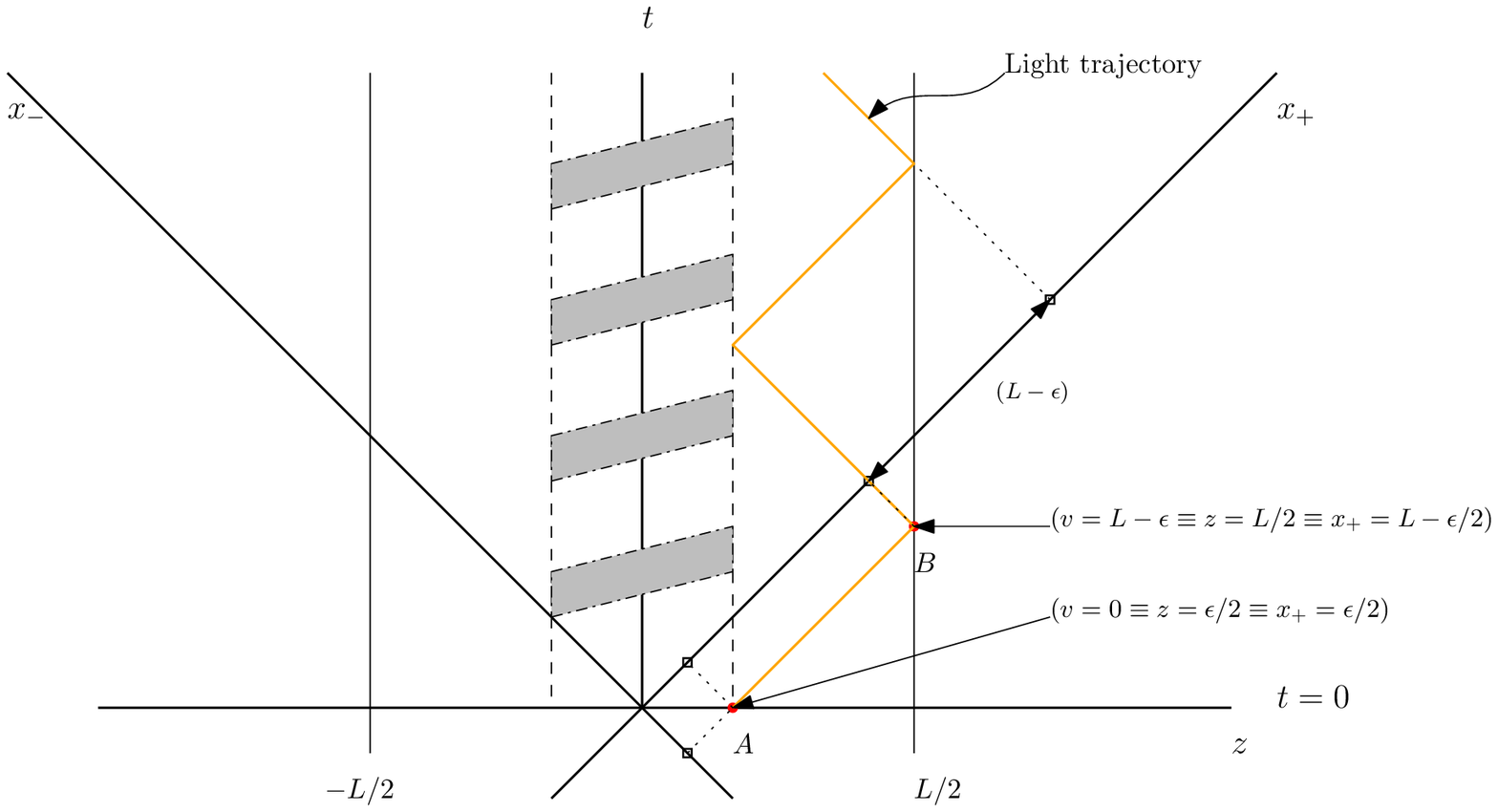}  
     \label{fig:fff}
    \end{center}
    Figure 3: diagram representation of both frames $\{t,z\}$ and $\{x_{+},x_{-}\}$ the yellow line show the null lines of massless particles in the $\{t,z\}$ frame. The periodicity of the expression (\ref{eq:93}) by $L-\epsilon$ is represented on the $x_{+}$ axis, which corresponds also to the coordinate points A, B at which the massless particles get reflected.
 
 \begin{equation}
 \begin{split}
     & \bigg[w^2\sum_{n=1}^{\infty}\cos(\frac{2n\pi}{L-\epsilon}(x_{\pm}-\frac{\epsilon}{2}))  +m^2\sum_{n=1}^{\infty}\frac{\cos(\frac{2n\pi}{L-\epsilon}(x_{\pm}-\frac{\epsilon}{2}))}{n^2 -(m/w)^2} \bigg]^2=\\
     & \Bigg[ -\frac{w^2}{2}+\frac{w^2(L-\epsilon)}{2}\delta(v) -\frac{w m\pi}{2}\cos\bigg(\frac{m\pi}{2}(1-\frac{\epsilon}{L})-\frac{m\pi}{L}v\bigg)\frac{1}{\sin(\frac{m\pi}{2}(1-\frac{\epsilon}{L}))} + \frac{w^2}{2}\Bigg]^2 = \\
    & \frac{w^4(L-\epsilon)^2}{4}\delta^2(v) +\frac{w^2m^2\pi^2}{4}\frac{\cos^2\bigg(\frac{m\pi}{2}(1-\frac{\epsilon}{L})-\frac{m\pi}{L}v\bigg)}{\sin^2\bigg(\frac{m\pi}{2}(1-\frac{\epsilon}{L})\bigg)} -\frac{m\pi w^3(L-\epsilon)}{2}\delta(0)\frac{\cos(\frac{m\pi}{2}(1-\frac{\epsilon}{L}))}{\sin(\frac{m\pi}{2}(1-\frac{\epsilon}{L}))}\\
 \end{split}
 \end{equation}
 From this and equation (\ref{eq:93}), compute the energy density for $v=0$, for which we get

\begin{equation}\label{eq:97}
    \begin{split}
        \langle 0_g| T_{\pm\pm} |0_g\rangle =
        & \sum_{\stackrel{{}_{\mathrm{odd}}}{m = 1}}^{\infty}\cos^2(\frac{m\pi}{2}\frac{\epsilon}{L})\bigg[\frac{2}{m\pi}\delta^2(0)+\frac{m\pi}{2L^2} \frac{\sin^2(\frac{m\pi}{2}\frac{\epsilon}{L})}{\cos^2(\frac{m\pi}{2}\frac{\epsilon}{L})}-\frac{2}{L}\delta(0)\cot\bigg(\frac{m\pi}{2}(1-\frac{\epsilon}{L})\bigg)\bigg] + \\
        & \sum_{\stackrel{{}_{\mathrm{even}}}{m = 2}}^{\infty}\sin^2(\frac{m\pi}{2}\frac{\epsilon}{L})\bigg[ \frac{2}{m\pi}\delta^2(0)+\frac{m\pi}{2L^2} \frac{\sin^2(\frac{m\pi}{2}\frac{\epsilon}{L})}{\cos^2(\frac{m\pi}{2}\frac{\epsilon}{L})} -\frac{2}{L}\delta(0)\cot\bigg(\frac{m\pi}{2}(1-\frac{\epsilon}{L})\bigg)\bigg]
    \end{split}
\end{equation}  \\
  
\begin{multicols}{2}   
  Similarly the sums in the second and the forth lines are computed via

\begin{equation}\label{eq:98}
    \sum_{k=1}^{\infty}\frac{k\sin(ky)}{k^2-a^2}=\frac{\pi}{2}\sin[a(\pi-y)]\csc(a\pi)
\end{equation}
valid for valid for $0< y < 2\pi$, then they become 
\begin{center}
\noindent\rule{8cm}{0.4pt} 
\end{center}

\begin{equation}\label{eq:99}
     \bigg[ \sum_{n=1}^{\infty}\frac{n\sin(\frac{2n\pi}{L-\epsilon}(x_{\pm}-\frac{\epsilon}{2}))}{n^2 -(m/w)^2} \bigg]^2= \frac{\pi^2}{4}\frac{\sin^2\bigg(\frac{m\pi}{2}(1-\frac{\epsilon}{L})-\frac{m\pi}{L}v\bigg)}{\sin^2\bigg(\frac{m\pi}{2}(1-\frac{\epsilon}{L})\bigg)}
\end{equation}

for $0<v <L-\epsilon $. 
Now, gathering all the pieces together, and generalizing for all $v$, we get 
\end{multicols}

\begin{equation}\label{eq:100}
\begin{split}
\langle 0_g| T_{\pm\pm} |0_g\rangle =
& \frac{1}{\pi}\sum_{\stackrel{{}_{\mathrm{odd}}}{m = 1}}^{\infty}\frac{1}{m}\Bigg[\delta^2(v-d(L-\epsilon))-\frac{m\pi}{L}\delta(0)\cot\big( \frac{m\pi}{2}(1-\frac{\epsilon}{L})\big)\Bigg]\cos^2(\frac{m\pi}{2}\frac{\epsilon}{L}) \\
& + \begin{cases}
      & \frac{\pi}{4L^2}\sum_{\stackrel{{}_{\mathrm{odd}}}{m = 1}}^{\infty}m\sin^2(\frac{m\pi}{2}\frac{\epsilon}{L})\text{\hspace{2.29cm}; $v=d(L-\epsilon)$} \\
      & \frac{\pi}{4L^2}\sum_{\stackrel{{}_{\mathrm{odd}}}{m = 1}}^{\infty}m \xrightarrow{\text{regularized}}\frac{\pi}{12L^2} \text{\hspace{1.55cm}; otherwise}\\ 
    \end{cases} \\
&+ \frac{1}{\pi}\sum_{\stackrel{{}_{\mathrm{even}}}{m = 2}}^{\infty}\frac{1}{m}\Bigg[\delta^2(v-d(L-\epsilon))-\frac{m\pi}{L}\delta(0)\cot\big( \frac{m\pi}{2}(1-\frac{\epsilon}{L})\big)\Bigg]\sin^2(\frac{m\pi}{2}\frac{\epsilon}{L}) \\
& + \begin{cases}
      & \frac{\pi}{4L^2}\sum_{\stackrel{{}_{\mathrm{even}}}{m=2}}^{\infty}m\tan^2(\frac{m\pi}{2}\frac{\epsilon}{L})\sin^2(\frac{m\pi}{2}\frac{\epsilon}{L})\text{\hspace{0.5cm}; $v=d(L-\epsilon)$} \\
      & \frac{\pi}{4L^2}\sum_{\stackrel{{}_{\mathrm{even}}}{m=2}}^{\infty}m\tan^2(\frac{m\pi}{2}\frac{\epsilon}{L})\text{\hspace{2.2cm}; otherwise} \\      
    \end{cases} \\
\end{split}
\end{equation}
\vspace{1mm}
\begin{multicols}{2}
We notice that we ended up with different terms for the energy density, more particularly $\delta$ and $\delta^2$ terms and trigonometric terms that one could try to regularize. Additionally, we can look at the number of created particles, let's call it ${\mathcal{N}}$, it is expressed as follows
\begin{equation}\label{eq..5.39}
    \begin{split}
     & {\mathcal{N}}_n(\epsilon/L) = \langle 0_g|a_{n}^{(l)\dagger}a_{n}^{(l)}|0_g\rangle
     = \sum_{m=1}^{\infty}|\beta_{mn}|^2\\
     &=\sum_{m(odd)=1}^{\infty}|\beta_{mn}|^2+\sum_{m(even)=2}^{\infty}|\beta_{mn}|^2\\
     &=\frac{4n}{\pi^2}\sum_{m(odd)=1}^{\infty}\frac{1}{m[2n+m(1-\epsilon/L)]^2}\cos^2(\frac{m\pi}{2}\frac{\epsilon}{L})\\
     &+ \frac{4n}{\pi^2}\sum_{m(even)=2}^{\infty}\frac{1}{m[2n+m(1-\epsilon/L)]^2}\sin^2(\frac{m\pi}{2}\frac{\epsilon}{L})\\
     &=\frac{4n}{\pi^2}\sum_{m(odd)=1}^{\infty}\frac{1}{m[2n+m(1-\epsilon/L)]^2}\\
     &-\frac{4n}{\pi^2}\sum_{m=1}^{\infty}\frac{1}{m[2n+m(1-\epsilon/L)]^2}\sin^2(\frac{m\pi}{2}\frac{\epsilon}{L})
    \end{split}
    \end{equation}
    The sums with respect to $m$ are convergent however if we sum with respect to $n$ the quantity $ {\mathcal{N}}_n(\epsilon/L)$, which is depicted in Fig.4, blows up, this result is in accordance with the energy density calculated previously. We see from the diagram that the number density is inversely proportional to the size of the thick Dirichlet wall.
    \includegraphics[width=8cm]{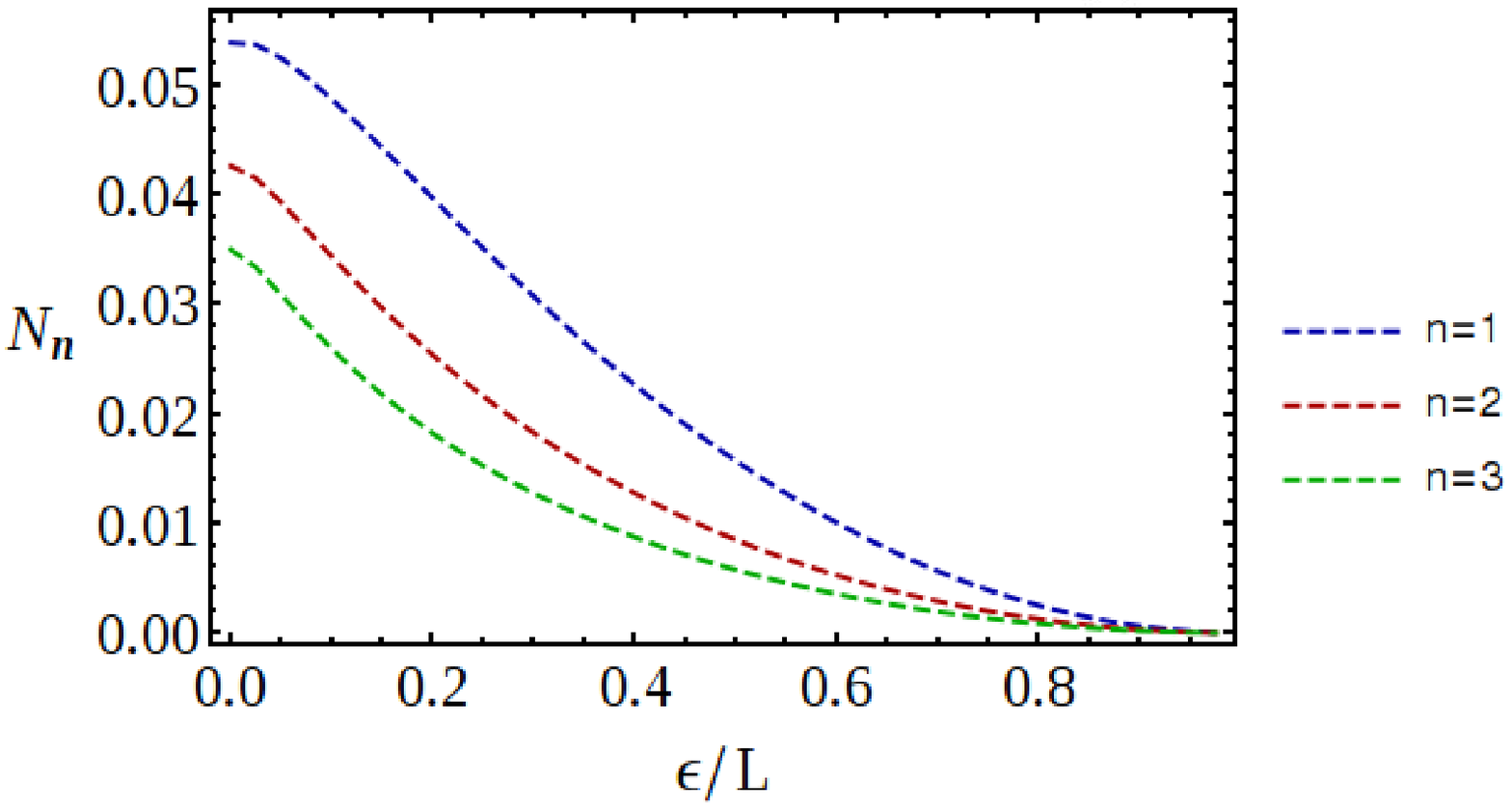}
    \label{fig-particles}
    Figure 4\ref{fig-particles}: diagram representation of the change of particle creation inside a cavity depending on the ratio $\epsilon/L$. This is done for the modes $n=1$, $n=2$, and $n=3$ (blue, reg, green, respectively)\\

  \subsection{Case 2: $\epsilon/L$ rational}
  We have $\epsilon/L=1-2n/m$ , meaning, it could be written as $\epsilon/L = \frac{p}{q}=\alpha$, with $p$ and $q$ integers. Now, since $\epsilon/L$ is a non-negative quantity, then $m>2n$. This implies that for $m$ odd, $1\leq n<\floor*{\frac{m}{2}}$ and for $m$ even, $1\leq n\leq\frac{m}{2}$. Similarly to the previous case, the quantity is $\langle 0_g| T_{\pm\pm} |0_g\rangle$ for $t>0$, hence, here equation (\ref{eq:92}) becomes:
  \begin{center}
\noindent\rule{8cm}{0.4pt} 
\end{center}
 \begin{equation}\label{eq.case2}
    \begin{split}
    &\langle 0_g| T_{\pm\pm} |0_g\rangle\\
    &=\sum_{l=1}^{2}\bigg(\sum_{\stackrel{{}_{\mathrm{odd}}}{m = 3}}^{\infty}\sum_{n=1}^{\floor*{\frac{m}{2}} }\sum_{n'=1}^{\floor*{\frac{m}{2}}}[(\alpha_{mn}^{l}\alpha_{mn'}^{l})\Re(\partial_{\pm}h_{n}^{l}\partial_{\pm}h_{n'}^{l*})]\\
    &+\sum_{\stackrel{{}_{\mathrm{even}}}{m = 2}}^{\infty}\sum_{n=1}^{\frac{m}{2}}\sum_{n'=1}^{\frac{m}{2}}[(\alpha_{mn}^{l}\alpha_{mn'}^{l})\Re(\partial_{\pm}h_{n}^{l}\partial_{\pm}h_{n'}^{l*})]\bigg)
    \end{split}
 \end{equation}

  Replacing the corresponding Bogoliobuv coefficients from (\ref{eq:90}) in (\ref{eq.case2}), we get the expression (\ref{eq-2case}) below
\end{multicols}  
\begin{equation}\label{eq-2case}
    \begin{split}
    &\langle 0_g| T_{\pm\pm} |0_g\rangle\\
    &=\sum_{l=1}^{2}\bigg(\sum_{\stackrel{{}_{\mathrm{odd}}}{m = 3}}^{\infty}\sum_{n=1}^{\floor*{\frac{m}{2}} }\sum_{n'=1}^{\floor*{\frac{m}{2}}}\bigg[\frac{\sqrt{nn'}}{m}\sin((m-2n)\frac{\pi}{2})\sin((m-2n')\frac{\pi}{2})\Re(\partial_{\pm}h_{n}^{l}\partial_{\pm}h_{n'}^{l*})\bigg]\\
    &+\sum_{\stackrel{{}_{\mathrm{even}}}{m = 2}}^{\infty}\sum_{n=1}^{\frac{m}{2}}\sum_{n'=1}^{\frac{m}{2}}\bigg[\frac{\sqrt{nn'}}{m}\cos((m-2n)\frac{\pi}{2})\cos((m-2n')\frac{\pi}{2})\Re(\partial_{\pm}h_{n}^{l}\partial_{\pm}h_{n'}^{l*})\bigg]\bigg)\\
    &=\sum_{\stackrel{{}_{\mathrm{odd}}}{m = 3}}^{\infty}\sum_{n=1}^{\floor*{\frac{m}{2}} }\sum_{n'=1}^{\floor*{\frac{m}{2}}}\frac{1}{2\pi m}\bigg[p_n \sin((m-2n)\frac{\pi}{2})\cos(p_n(x_{\pm} - \frac{\epsilon}{2}))\bigg(n\to n'\bigg)-\\
    &p_{n} \sin((m-2n)\frac{\pi}{2})\sin(p_n(x_{\pm} - \frac{\epsilon}{2}))\bigg(n\to n'\bigg)  \bigg]\\
    &+\sum_{\stackrel{{}_{\mathrm{even}}}{m = 2}}^{\infty}\sum_{n=1}^{\frac{m}{2}}\sum_{n'=1}^{\frac{m}{2}}\frac{1}{2\pi m}\bigg[p_n \cos((m-2n)\frac{\pi}{2})\cos(p_n(x_{\pm} - \frac{\epsilon}{2}))\bigg(n\to n'\bigg) -\\
    &p_{n} \cos((m-2n)\frac{\pi}{2})\sin(p_n(x_{\pm} - \frac{\epsilon}{2}))\bigg(n\to n'\bigg)   \bigg]\\
    \end{split}
\end{equation}

\begin{multicols}{2}
  
 Manipulating the expression (\ref{eq-2case}) mathematically yields the following
  
 \begin{equation}\label{2ndEnrgy}
    \begin{split}
    &\langle 0_g| T_{\pm\pm} |0_g\rangle\\
    &=\sum_{\stackrel{{}_{\mathrm{odd}}}{m = 3}}^{\infty}\frac{1}{2\pi m}\frac{4\pi^2}{L^2(1-\alpha^2)}\bigg(\bigg[\sum_{n=1}^{\floor*{\frac{m}{2}} }n (-1)^{n}\cos(ny)\bigg]^2\\
    &- \bigg[\sum_{n=1}^{\floor*{\frac{m}{2}} }n  (-1)^{n}\sin(ny) \bigg]^2\bigg)\\
    &+\sum_{\stackrel{{}_{\mathrm{even}}}{m = 2}}^{\infty}\frac{1}{2\pi m}\frac{4\pi^2}{L^2(1-\alpha^2)}\bigg(\bigg[\sum_{n=1}^{\frac{m}{2} }n (-1)^{n}\cos(ny)\bigg]^2\\
    &-\bigg[\sum_{n=1}^{\frac{m}{2} }n  (-1)^{n}\sin(ny) \bigg]^2\bigg)
    \end{split}
\end{equation} 

with $y=\frac{2\pi}{L(1-\alpha)}(x_{\pm}-\alpha\frac{L}{2})$. To simplify the writing, we define
    \begin{equation}\label{eqFc}
    \begin{split}
     f_c(m)
     &=\sum_{n=1}^{m}n (-1)^{n}\cos(ny)\\
    \end{split}
    \end{equation}
\hspace{5mm}and
    \begin{equation}\label{eqFs}
    \begin{split}
     f_s(m)
     &=\sum_{n=1}^{m}n (-1)^{n}\sin(ny)\\
    \end{split}
    \end{equation}
Finally, the sums in (\ref{2ndEnrgy}) could be redefined in such a way to end up with the following expression for the energy density
\begin{equation}\label{ENRG}
    \begin{split}
    &\langle 0_g| T_{\pm\pm} |0_g\rangle\\
    &=\frac{2\pi}{L^2(1-\alpha^2)}\sum_{m = 1}^{\infty}\frac{4m+1}{2m(m+1)}\bigg(\bigg[f_c(m)\bigg]^2- \\
    &\bigg[f_s(m) \bigg]^2\bigg)\\
    \end{split}
\end{equation}\\

Here, the sums are too complicated, see \ref{AppendixC} for details. Hence, to have a feeling on what the energy would be, we consider its quantity at one point of space i.e. we fix, say, $x_+$ for some specific value of $\epsilon/L$. We choose $x_+$ in such a way (\ref{ENRG}) becomes as simple as possible, for example $y$ is set to zero. Hence, from (\ref{sumC}) in appendix \ref{AppendixC}, we get
\begin{equation}\label{final-energy}
    \begin{split}
    &\langle 0_g| T_{\pm\pm} |0_g\rangle\\
    &=\frac{2\pi}{L^2(1-\alpha^2)}\sum_{m = 1}^{\infty}\frac{4m+1}{2m(m+1)}\frac{1}{16}\bigg(\\
    &(2m+1)^2 -2(2m+1)(-1)^m + 1\bigg)\\
    \end{split}
\end{equation}\\
or
\begin{equation}\label{final-energy}
    \begin{split}
    &\langle 0_g| T_{\pm\pm} |0_g\rangle\\
    &=\frac{\pi}{8L^2(1-\alpha^2)}\sum_{m = 1}^{\infty}\frac{4m+1}{m(m+1)}\bigg(\\
    &2m^2 +2m -2m(-1)^m - (-1)^m + 1\bigg)\\
    \end{split}
\end{equation}\\
Here, $(-1)^m$ is the only converging term, namely, $\sum_{m = 1}^{\infty}\frac{4m+1}{m(m+1)}(-1)^m=\frac{1}{2}(log(4)-3)$. The other terms need to be regularized. Using the Riemann-zeta regularization form (i.e. $\zeta(s)=\sum_{n = 1}^{\infty}\frac{1}{n^s}$), the Euler sum of $\sum_{n=1}^{\infty}n=-1/12$ and the fact that $\sum_{n=1}^{\infty}1=-1/2$ (this is computed via the \emph{analytic continuation} of the zeta-regularization check \cite{Master}), the sum $\sum_{m=1}^{\infty}\frac{4m+1}{m(m+1)}(2m^2 +2m + 1)$ is regularized to be: $-\frac{14}{3}+4\zeta(1)$\\
\hspace{5mm}On the Other hand the remaining sum,
\\$\sum_{m=1}^{\infty}\frac{4m+1}{m(m+1)}(-2m(-1)^m)$ , is simply given by: $6\log(2)+2$; Finally, gathering all the terms together, one finds as a result the following:
\begin{equation}\label{final-energy}
    \begin{split}
    &\langle 0_g| T_{\pm\pm} |0_g\rangle\\
    &=\frac{\pi}{8L^2(1-\alpha^2)}\bigg(-\frac{7}{6}+4\zeta(1)+5\log(2)\bigg)\\
    \end{split}
\end{equation}\\
In this equation, we still have a diverging term, not in the Dirac $\delta$ form but in the Riemann-zeta form: $\zeta(1)$.
\section{Conclusion and discussion}
One of the dynamical effects that gives rise to the creation of particles is the sudden appearance of a wall in a Casimir cavity. We present a material in which we discuss the case of thick Dirichlet wall. We follow the approach of \cite{Artcl-2016} which initiated a similar investigation for the case of an infinitesimally thin Dirichlet wall that appears (and disappears) suddenly in a cavity. We have been able to obtain expressions for the energy which appears to be dependent on the thickness of the wall, being an irrational or a rational number in terms of the cavity size. We demonstrate that in the configuration of the thick wall, the expression for the energy density, especially in the \emph{irrational} thickness case, becomes quite involved, Eq. (\ref{eq:93}), so that it requires a careful normalization of ultraviolet divergences associated with the particle creation in the dynamical Casimir setup. The complexity of the energy density could be attributed to the type of the boundary and to irrational length space in the cavity which makes the field modes more disturbed, this result could be seen as a direct complex version of the result derived in \cite{Artcl-2016}, where terms such as $\delta^2$ are also obtained. However, surprisingly in the case of the \emph{rational} size, energy density was reduced to only a set of sums and no $\delta^2$ terms included, the new term that has appeared instead was the $\zeta(1)$. Moreover, contrarily to the first situation (and to the results of \cite{Artcl-2016} for that matter), here there was no distinction between the null lines ($\{x_+,x_-\}$) and the other points of space inside the cavity.  
\section*{Acknowledgements}
I am particularly grateful for the assistance given by Dr. Maxim Chernodub who was my supervisor in my master's project, from which this article is inspired.
\end{multicols}

\vspace{2cm}

\appendix
\section{Distinction between ($\epsilon/L = 1-2n/m$) and ($\epsilon/L \neq 1-2n/m$)}
\label{AppendixA}
Let's compute the first two terms in equation (\ref{eq:90}), therefore, for the case $m$ being odd and using the definition of $\alpha_{mn}^{l}$, we have 
\begin{equation}
    \begin{split}
        \alpha_{mn}^{1}
        &=i\int_{\epsilon/2}^{L/2}dz\bigg[\frac{1}{\sqrt{n\pi}}e^{ip_{n}t}\sin(p_{n}(z-\frac{\epsilon}{2}))\frac{(-iq_{m})}{\sqrt{m\pi}}e^{-iq_{m}t}\cos(q_{m}z)\\
        &-\frac{ip_{n}}{\sqrt{n\pi}}e^{ip_{n}t}\sin(p_{n}(z-\frac{\epsilon}{2}))\frac{1}{\sqrt{m\pi}}e^{-iq_{m}t}\cos(q_{m}z)\bigg]\bigg|_{t=0}\\
        &=\int_{\epsilon/2}^{L/2}dz\frac{1}{\pi\sqrt{mn}}(p_n+q_m)\sin(p_{n}(z-\frac{\epsilon}{2}))\cos(q_{m}z)\\
        &=\frac{1}{2\pi\sqrt{mn}}(p_n+q_m)\int_{\epsilon/2}^{L/2}dz\bigg[\sin\bigg((p_n+q_m)z-p_n\frac{\epsilon}{2}\bigg)+\sin\bigg((p_n-q_m)z-p_n\frac{\epsilon}{2}\bigg)\bigg]\\
        &=-\frac{1}{2\pi\sqrt{mn}}\cos\bigg((p_n+q_m)z-p_n\frac{\epsilon}{2}\bigg)\bigg|_{\epsilon/2}^{L/2}-\frac{1}{2\pi\sqrt{mn}}\frac{p_n+q_m}{p_n-q_m}\cos\bigg((p_n-q_m)z-p_n\frac{\epsilon}{2}\bigg)\bigg|_{\epsilon/2}^{L/2}
    \end{split}
\end{equation}
Here there is a subtlety about the ration in the second term of the last equality. Two situations are present here; the first is when $p_n\neq q_m$, the second is when $p_n = q_m$, recall that $p_n=\frac{2n\pi}{L-\epsilon}$ and $q_m=\frac{m\pi}{L}$. Hence,\\

for $p_n\neq q_m$ or, equivalently, $\frac{\epsilon}{L}\neq 1-\frac{2n}{m}$, we have
\begin{equation}
      \alpha_{mn}^{1}=\frac{1}{\pi\sqrt{mn}}\frac{2n}{2n-m(1-\epsilon/L)}\cos\Big(\frac{m\pi}{2}\frac{\epsilon}{L}\Big)
\end{equation}

and for $p_n = q_m$ or, equivalently, $\frac{\epsilon}{L}= 1-\frac{2n}{m}$, with $m\geq2n$ (because $\epsilon/L$ is positive), the computation of the first term gives zero, while the second gives 
\begin{equation}
    \alpha_{mn}^{1}=-\sqrt{\frac{n}{m}}\sin\Big((m-2n)\frac{\pi}{2}\Big)
\end{equation}
where in the computation of this second term, the following property (Hospital's rule) is used
\begin{equation}
    \lim_{x\to a}\frac{\cos(x)}{x-a}=\lim_{x\to a}-\sin(x) \equiv -\sin(a)\,,
\end{equation}
where $a$ makes cosine vanishing, $\cos a = 0$.
The case for $l=2$ as well as the other coefficients are computed similarly.\\

\section{The sums in (\ref{eqFc}) and (\ref{eqFs})}
\label{AppendixC}
We have
\begin{equation}
    \begin{split}
    f_c(m)
    &=\sum_{n=1}^{m}n (-1)^{n}\cos(ny)\\
    &=\frac{1}{4}\bigg(-\sec^2(y/2)\cos(my+\pi m +y)+2m\sec(y/2)\cos(\frac{1}{2}(2my+2\pi m + y))\\
    &+2\sec(y/2)\cos(\frac{1}{2}(2my+2\pi m + y))+\cos(y)\sec^2(y/2)+2\cos(\frac{1}{2}(y-2\pi))\sec(y/2)
    \bigg)
     \end{split}
\end{equation}
and
\begin{equation}
    \begin{split}
    f_s(m)
    &=\sum_{n=1}^{m}n (-1)^{n}\cos(ny)\\
    &=\frac{1}{4}\bigg(-\sec^2(y/2)\sin(my+\pi m +y)+2m\sec(y/2)\sin(\frac{1}{2}(2my+2\pi m + y))\\
    &+2\sec(y/2)\sin(\frac{1}{2}(2my+2\pi m + y))+\sin(y)\sec^2(y/2)+2\sin(\frac{1}{2}(y-2\pi))\sec(y/2)
    \bigg)
     \end{split}
\end{equation}
From which we can write
\begin{equation}\label{sumC}
    \begin{split}
    f_c(m)^2-f_s(m)^2
    &=\frac{1}{16}\bigg(\sin^2(my)[4(m+2)^2\beta^2-(\beta^2-1+2(m+1)(1+\beta))^2]\\
    &+\cos^2(my)[(2m+1+\beta^2)^2-4\beta^2]\\
    &+\sin(my)\cos(my)[4(m+2)(2m+1+\beta^2) +4(\beta^2-1+2(m+1)(1+\beta))\beta]\\
    &-4(m+2)\beta(1+\beta^2)(-1)^m\sin(my)\\
    &-2(2m+1+\beta^2)(1+\beta^2)(-1)^m\cos(my)\\
    &+(1+\beta^2)^2
    \bigg)
     \end{split}
\end{equation}
with $\beta=\tan(y/2)$.

\begin{center}
\noindent\rule{16cm}{0.4pt} 
\end{center}

\setlength{\columnsep}{1cm}

\end{document}